%% file: draftJournalPaper_byparts_v4.tex
\documentclass[aoas]{imsart}

\usepackage[utf8]{inputenc}
\usepackage{geometry}
\usepackage{rotating}
\usepackage{amsmath}
\usepackage{comment}
\usepackage{amsfonts}
\usepackage{url}
\usepackage{comment}
\usepackage{algorithm}
\usepackage{algpseudocode}
\usepackage{natbib}
\usepackage{enumitem}
\usepackage{color}

\algnewcommand\algorithmicinput{\textbf{Input:}}
\algnewcommand\Input{\item[\algorithmicinput]}
\algnewcommand\algorithmicoutput{\textbf{Output:}}
\algnewcommand\Output{\item[\algorithmicoutput]}

\startlocaldefs
\endlocaldefs

\begin{document}

\begin{frontmatter}

\title{Providing Accurate Models across Private Partitioned Data: Secure Maximum Likelihood Estimation}
\runtitle{Secure Maximum Likelihood Estimation}
\thankstext{T1}{Footnote to the title with the `thankstext' command.}


\begin{aug}
\author{\fnms{Joshua} \snm{Snoke}\ead[label=e1]{snoke@psu.edu}},
\author{\fnms{Timothy R.} \snm{Brick}\ead[label=e2]{tbrick@psu.edu}},
\author{\fnms{Aleksandra} \snm{Slavkovi\'{c}}\ead[label=e3]{abs12@psu.edu}}
\and
\author{\fnms{Michael D.} \snm{Hunter}\ead[label=e4]{mhunter1@ouhsc.edu}}

\runauthor{Snoke et. al}

\affiliation{Pennsylvania State University and University of Oklahoma Health Sciences Center}

\address{Statistics Department,\\ 
          \printead{e1,e3}}
\address{Human Development and Family Studies Department,\\ 
          \printead{e2}}
\address{Department of Pediatrics,\\ 
          \printead{e4}}

\end{aug}

\begin{abstract}
This paper focuses on the privacy paradigm of providing access to researchers to remotely carry out analyses on sensitive data stored behind firewalls. We address the situation where the analysis demands data from multiple physically separate databases which cannot be combined. Motivating this problem are analyses using multiple data sources that currently are only possible through extension work creating a trusted user network. We develop and demonstrate a method for accurate calculation of the multivariate normal likelihood equation, for a set of parameters given the partitioned data, which can then be maximized to obtain estimates. These estimates are achieved without sharing any data or any true intermediate statistics of the data across firewalls. We show that under a certain set of assumptions our method for estimation across these partitions achieves identical results as estimation with the full data. Privacy is maintained by adding noise at each partition. This ensures each party receives noisy statistics, such that the noise cannot be removed until the last step to obtain a single value, the true total log-likelihood. Potential applications include all methods utilizing parameter estimation through maximizing the multivariate normal likelihood equation. We give detailed algorithms, along with available software, and both a real data example and simulations estimating structural equation models (SEMs) with partitioned data.
\end{abstract}



\end{frontmatter}

\section{Introduction}
\label{sec:intro}
\input{introduction}

\section{Previous secure methods}
\label{sec:prevmeth}
\input{previous-secure-methods}

\section{Distributed likelihood estimation}
\label{sec:dle}
\input{distributed-likelihood-estimation}


\section{Secure algorithm for vertically partitioned data}
\label{sec:sharing}
\input{secure-algorithm}

\section{Real Data Example}
\label{sec:realdata}
\input{real-data-example}


\section{Simulations for Accuracy and Computational Complexity}
\label{sec:sims}
\input{simulations}

\section{Discussion}
\label{sec:discussion}
\input{discussion}

\bibliographystyle{imsart-nameyear}

\bibliography{../firewall}

\newpage
\section{Appendix: Internal algorithms}
\label{sec:app1}
Reference Notation:
\begin{itemize}
  \item $P_k$, $R_k$, $Q_k$, $M_k$ random noise matrices with dimensions $\mathbb{R}^{n \times p_k}, \mathbb{R}^{n \times p_k}, \mathbb{R}^{p_k \times n}, \mathbb{R}^{n \times p_k}$ respectively
  \item $A_k^1 = \Sigma_{x_kx_k|x_k^-}^{-1}(x_k - \tilde\mu_{x_k|x_k^-} + R_k)$
  \item $A_k^2 = \Sigma_{x_kx_k|x_k^-}^{-1}(x_k - \tilde\mu_{x_k|x_k^-} - R_k) + Q_k$
  \item $B_k = \tilde\mu_{x_k^+|x_k^-} + \Sigma_{x_kx_k^+|x_k^-}A^1_k$
  \item $C_k = \Sigma_{x_kx_k^+|x_k^-}(\Sigma_{x_kx_k|x_k^-})^{-1}$
  \item $\tilde\mu_{x_{k}^+|x_{k+1}^-} = B_{k} - M_k - C_{k}(R_{k} - P_{k})$
\end{itemize}

\begin{algorithm}
\caption{Central Node Initiate (\emph{CN\_Initiate})}
\label{alg:centralInit}
\begin{algorithmic}[1]
\Input $\mu, \Sigma$
\Output $\Sigma_{x_kx_k|x_k^-}, P_k, \tilde\mu_{x_k}, C_k$ $\forall k \in K$
\For{k in 1,..., K}
  \State Compute $\Sigma_{x_kx_k|x_k^-}$
  \State Generate random $P_k \in \mathbb{R}^{n \times p_k}$
  \State Compute $\tilde\mu_{x_k} = \mu_{x_k} + P_k$
  \State Compute $C_k = \Sigma_{x_kx_k^+|x_k^-}(\Sigma_{x_kx_k|x_k^-})^{-1}$
\EndFor
\end{algorithmic}
\end{algorithm}

\begin{algorithm}
\caption{External Node Computation (\emph{EN\_Compute})}
\label{alg:externalComp}
\begin{algorithmic}[1]
\Input $\tilde\mu_{x_k|x_k^-}, \Sigma_{x_kx_k|x_k^-}, X_k, \tilde LL_{\Sigma_{j = 1}^{k-1} j}^*$
\Output $\tilde LL_{\Sigma_{j = 1}^{k} j}, A_k^1, A_k^2, R_k, Q_k$
\State Generate random $R_k \in \mathbb{R}^{n \times p_k}$
\State Generate random $Q_k \in \mathbb{R}^{p_k \times n}$
\State Compute $A_k^1 = \Sigma^{-1}_{x_kx_k|x_k^-}(X_k - \tilde\mu_{x_k|x_k^-} + R_k)$
\State Compute $A_k^2 = \Sigma^{-1}_{x_kx_k|x_k^-}(X_k - \tilde\mu_{x_k|x_k^-} - R_k) + Q_k$
\State Compute $\tilde LL_{k} = computeNoisyLL(\tilde\mu_{x_k|x_k^-}, \Sigma_{x_kx_k|x_k^-}, X_k, R_k)$
\If{$\tilde LL_{\Sigma_{j = 1}^{k-1} j}^* != \emptyset$}
  \State $\tilde LL_{\Sigma_{j = 1}^{k} j} = \tilde LL_{\Sigma_{j = 1}^{k-1} j}^* + \tilde LL_{k}$
\Else
  \State $\tilde LL_{\Sigma_{j = 1}^{k} j} = \tilde LL_{k}$
\EndIf
\end{algorithmic}
\end{algorithm}

\begin{algorithm}
\caption{Compute Noisy LL (\emph{computeNoisyLL})}
\label{alg:computeLL}
\begin{algorithmic}[1]
\Input $\tilde\mu_{x_k|x_k^-}, \Sigma_{x_kx_k|x_k^-}, X_k, R_k$
\Output $\tilde LL_{k}$
\State Compute $\tilde LL_{k} = \Sigma^n_{i = 1} [p_klog(2\pi) + log(|\Sigma_{x_kx_k|x_k^-}|) + (x_{ki} - \tilde\mu_{x_k|x_k^-} + R_k)\Sigma_{x_kx_k}^{-1}(x_{ki} - \tilde\mu_{x_k|x_k^-} - R_k)^T + R_k^T\Sigma_{x_kx_k|x_k^-}^{-1}R_k] = LL_{k} - A^1_kP_k^T - P_k(A^2_k)^T - P_k\Sigma_{x_kx_k}^{-1}P_k^T + P_kQ_k$
\end{algorithmic}
\end{algorithm}

\begin{algorithm}
\caption{Central Node Adjustment (\emph{CN\_Adjust})}
\label{alg:centralAdjust}
\begin{algorithmic}[1]
\Input $A_k^1, \tilde\mu_{x_k^+|x_k^-}, \Sigma_{x_kx_k|x_k^-}$
\Output $B_k$
\State Compute $B_k = \tilde\mu_{x_k^+|x_k^-} + \Sigma_{x_kx_k^+|x_k^-}A_k^1$
\end{algorithmic}
\end{algorithm}

\begin{algorithm}
\caption{External Node Adjustment (\emph{EN\_Adjust})}
\label{alg:externalAdjust}
\begin{algorithmic}[1]
\Input $B_k, C_k, R_k, P_k, Q_k, \tilde LL_{\Sigma_{j = 1}^{k} j}, M_k$
\Output $\tilde LL_{\Sigma_{j = 1}^{k} j}^*, \tilde\mu_{x_k^+|x_{k+1}^-}, \tilde\mu^*_{x_{k+1}^+|x_{k+1}^-}, M_{k+1}$
\State Compute $\tilde LL_{\Sigma_{j = 1}^{k} j}^* = \tilde LL_{\Sigma_{j = 1}^{k} j} - P_kQ_k^T$
\If{$M_k != \emptyset$}
  \State Compute $\tilde\mu_{x_k^+|x_{k+1}^-} = B_k - M_k - C_k(R_k - P_k)$
\Else
  \State Compute $\tilde\mu_{x_k^+|x_{k+1}^-} = B_k - C_k(R_k - P_k)$
\EndIf
\State Generate $M_{k+1} \in \mathbb{R}^{n \times p_k}$
\State Compute $\tilde\mu^*_{x_{k+1}^+|x_{k+1}^-} = \tilde\mu_{x_{k+1}^+|x_{k+1}^-} + M_{k+1}$
\end{algorithmic}
\end{algorithm}

\begin{algorithm}
\caption{First Node Adjustment (\emph{FN\_Adjust})}
\label{alg:firstAdjust}
\begin{algorithmic}[1]
\Input $\tilde LL_{\Sigma_{j = 1}^{K} j}, P_K, Q_K$
\Output $\tilde LL_{\Sigma_{j = 1}^{K} j}^*$
\State Compute $\tilde LL_{\Sigma_{j = 1}^{K} j}^* = \tilde LL_{\Sigma_{j = 1}^{K} j} - P_KQ_K^T$
\end{algorithmic}
\end{algorithm}

\begin{algorithm}
\caption{Central Node Final De-Noising (\emph{CN\_Final})}
\label{alg:centralFinal}
\begin{algorithmic}[1]
\Input $\tilde LL_{\Sigma_{j = 1}^{K} j}^*, P_k, A_k^1, A_k^2, \Sigma_{x_kx_k|x_k^-}$ $\forall k \in K$
\Output $LL_{\Sigma_{j = 1}^{K} j}$
\For{k in 1,..., K}
\State Compute $LL_{Noisek} = \Sigma^n_{i = 1} [A^1_kP_k^T + P_k(A^2_k)^T + P_k\Sigma_{x_kx_k|x_k^-}^{-1}P_k^T]$
\EndFor
\State Compute $LL_{\Sigma_{j = 1}^{K} j} = \tilde LL_{\Sigma_{j = 1}^{K} j}^* + \Sigma_{j = 1}^{K} LL_{Noisej}$
\end{algorithmic}
\end{algorithm}

\newpage
\section{Appendix: Non-Secure algorithm}
\label{sec:app2}
\begin{algorithm}
\caption{Non-Secure Passing Algorithm}
\label{alg:nontrusted}
\begin{algorithmic}[1]
\Input $\mu, \Sigma$ (cental node), $X_k \in \mathbb{R}^{n \times p_k}$ $\forall k \in K$ (data nodes)
\Output $LL_{\Sigma_{j = 1}^{K} j}$
\State CN $\to$ $DN_1$: $\Sigma$, $\mu$
\State $DN_1$ compute: $\Sigma_{x_1x_1}$, $\mu_{x_1}$, $\Sigma_{x_1^+x_1^+|x_1}$, $\hat\mu_{x_1^+|x_1}$
\State $DN_1$ compute: $LL_{1}$
\State $DN_1$ $\to$ $DN_2$: $LL_{1}$, $\Sigma_{x_1^+x_1^+|x_1}$, $\hat\mu_{x_1^+|x_1}$
\For{k in 2,...,(K-1)}
  \State $DN_k$ compute: $\Sigma_{x_kx_k|x_k^-}$, $\hat\mu_{x_k|x_k^-}$, $\Sigma_{x_k^+x_k^+|x_{k+1}^-}$, $\hat\mu_{x_k^+|x_{k+1}^-}$
  \State $DN_k$ compute: $LL_{k}$
  \State $DN_k$ $\to$ $DN_{k+1}$: $LL_{\Sigma_{j = 1}^{k} j}$, $\Sigma_{x_k^+x_k^+|x_{k+1}^-}$, $\hat\mu_{x_k^+|x_{k+1}^-}$
\EndFor
\State $DN_K$ compute: $\Sigma_{x_Kx_K|x_K^-}$, $\hat\mu_{x_K|x_K^-}$
\State $DN_K$ compute: $LL_{K}$
\State $DN_K$ $\to$ $CN$: $LL_{\Sigma_{j = 1}^{K} j}$
\end{algorithmic}
\end{algorithm}

\section{Appendix: Data leakage evaluation}
\label{sec:app3}
We acknowledge an asymmetry among the objects received by the different nodes. By showing that none of the nodes receive disclosive statistics, we make this asymmetry irrelevant.

\subsection{Node C}

\subsubsection{Starting objects}

Node C holds $\Sigma$ and $\mu$ (model defined parameters not statistics).

\subsubsection{Received objects}

\begin{itemize}
  \item $A_k^1$ for $k \in {1...K}$
  \item $A_k^2$ for $k \in {1...K}$
  \item $\tilde\mu_{x_{k}^+|x_{k}^-}^*$ for $k \in {2...K-1}$
  \item $\tilde LL_{\Sigma_{j = 1}^{K} j}^*$
\end{itemize}

\subsubsection{Anaylsis}

The central node recieves $A_k^1$ and $A_k^2$ for $i \in {1...K}$, such that:
\begin{align}
  \begin{split}
    A_k^1 = \Sigma_{x_kx_k|x_k^-}^{-1}(X_k - \tilde\mu_{x_k|x_k^-} + R_k) = \Sigma_{x_kx_k|x_k^-}^{-1}(X_k - \mu_{x_k|x_k^-} - P_k + R_k)\\
    A_k^2 = \Sigma_{x_kx_k|x_k^-}^{-1}(X_k - \tilde\mu_{x_k|x_k^-} - R_k) + Q_k = \Sigma_{x_kx_k|x_k^-}^{-1}(X_k - \mu_{x_k|x_k^-} - P_k - R_k) + Q_k
  \end{split}
\end{align}
Clearly it is important that $X_k$ should not be recovered, and for $k > 1$, $\tilde\mu_{x_k|x_k^-}$ is also risky because it is a statistic. Importantly, the central node knows $\Sigma_{x_kx_k|x_k^-}^{-1}$ and $P_k$, so these can be differenced out. To protect disclosure, $R_k$ and $Q_k$ are random noise added which the central node does not know. $R_k$ protects the values in $A_k^1$ and $Q_k$ protects from learning $R_k$ by differencing the two equations (both $A_1$ and $A_2$ are needed to calculate the correct log-likelihood).\\

Next, the central node receives $\tilde\mu_{x_{k}^+|x_{k - 1}}^*$ for $k \in {2...K-1}$ such that:
\begin{align}
  \begin{split}
    \tilde\mu_{x_{k}^+|x_{k}^-}^* = \tilde\mu_{x_{k}^+|x_{k}^-} + M_k = \mu_{x_{k}^+|x_{k}^-} + P_k + M_k
  \end{split}
\end{align}
The central node knows $P_k$, so $M_k$ is essential here to protect the true value of the conditional mean statistic. With that, there is no way for the central node to recover it.\\

Lastly, the central node receives $\tilde LL_{\Sigma_{j = 1}^{K} j}^*$. They can remove all the noise (as we want) to get the true total log-likelihood, ${LL_{\Sigma_{j = 1}^{K} j}}$, since this is one value composed across the entire set of partitioned databases. As assumed, this total is not risky and is necessary to obtain accurate estimates. 

\subsection{Node 1}

\subsubsection{Starting objects}

Node $1$ holds $X_1$

\subsubsection{Received objects}

\begin{itemize}
  \item $\Sigma_{x_1x_1}$
  \item $\tilde\mu_{x_1}$
  \item $P_K$
  \item $\tilde LL_{\Sigma_{j = 1}^{K} j}$
  \item $Q_K$
\end{itemize}

\subsubsection{Analysis}

The only object received by Node 1 that is conditioned on other nodes' data and potentially disclosive is $\tilde LL_{\Sigma_{j = 1}^{K} j}$.

Fortunately, this value is heavily perturbed with a variety of noise that Node 1 cannot access. Another note is that Node 1 could potentially estimate the value of $P_1$, since $\tilde\mu_{x_1} = \mu_{x_1} + P_1$ and the sample mean of Node 1's data should converge to $\mu_{x_1}$. Again though, this value is non-disclosive, since there isn't anything Node 1 can learn from having $P_1$.

\subsection{Nodes 2 through K}

\subsubsection{Starting objects}

Node $k$ holds $X_k$

\subsubsection{Received objects}

\begin{itemize}
  \item $\Sigma_{x_kx_k|k^-}$
  \item $B_{k-1}$
  \item $C_{k-1}$
  \item $P_{k-1}$
  \item $\tilde LL_{\Sigma_{j = 1}^{{k-1}} j}$
  \item $R_{k-1}$
  \item $Q_{k-1}$
  \item $M_{k-1}$ ($M_1 = 0$)
\end{itemize}

\subsubsection{Analysis}

There is an asymmetry among nodes, since Nodes 2 and following receive more information than Node 1. That being said, if none of it is actually disclosive the asymmetry is acceptable.\\

The first objects received that are potentially disclosive are $B_{k-1}$ and $C_{k-1}$ such that:
\begin{gather}
B_{k-1} = \tilde\mu_{x_{k-1}^+|x^-_{k-1}} + \Sigma_{x_{k-1}x_{k-1}^+|x^-_{k-1}}A^1_{k-1}
\end{gather}
\begin{gather*}
= \mu_{x_{k-1}^+|x^-_{k-1}} + (P_k \; P_{k+1} \; ... \; P_K) + \Sigma_{x_{k-1}x_{k-1}^+|x^-_{k-1}}(\Sigma_{x_{k-1}x_{k-1}|x^-_{k-1}})^{-1}(X_{k-1} - \tilde\mu_{x_{k-1}|x^-_{k-1}} - P_{k-1} + R_{k-1})
\end{gather*}
\begin{gather}
    C_{k-1} = \Sigma_{x_{k-1}x_{k-1}^+|x^-_{k-1}}(\Sigma_{x_{k-1}x_{k-1}|x^-_{k-1}})^{-1}
\end{gather}

Node $k$ is able to remove some of the noise, since:
\begin{gather*}
B_{k-1} - C_{k-1}(R_{k-1} - P_{k-1}) = \tilde\mu_{x_{k-1}^+} + \Sigma_{x_{k-1}x_{k-1}^+|x^-_{k-1}}\Sigma_{x_{k-1}x_{k-1}|x^-_{k-1}}^{-1}(X_{k-1} - \tilde\mu_{x_{k-1}|x^-_{k-1}}).
\end{gather*}
The key is that the noise $(P_k \; P_{k+1} \; ... \; P_K)$ is still present to protect the true value of the data and parameters. More importantly, since Node $k$ does not know any of the other individual components (apart from $P_{k-1}$ and $R_{k-1}$), it cannot decompose $B_{k-1}$ or $C_{k-1}$ futher.

$C_{k-1}$ is unperturbed, but two things convince us it is okay to share these. First it is a product of two matrices, neither of which Node $k$ knows. Second these are low risk objects. It is possible Node $k$ learns the scale of Node $(k-1)$'s data, but for now we consider that acceptable.

Lastly Node $k$ receives $\tilde LL_{\Sigma_{j = 1}^{k-1} j}$, a noisy version of the running total log-likelihood. Node $k$ can remove some noise based on $P_{k-1}$ and $Q_{k-1}$ (which we want), but does not know $A^1_{k-1}$ or $A^2_{k-1}$ and thus cannot recover the true $LL$ value.

\end{document}

%% file: introduction.tex
In many real-life settings, researchers wish to utilize data from separate databases but are unable to physically combine the data due to restrictions such as privacy concerns, proprietary issues, sheer size of the data, or massively distributed data such as proposed by \cite{boker2015maintained}. Consider situations such as in health where a researcher wishing to carry out a longitudinal study on PTSD for veterans who utilize different hospitals and primary care physicians, in education with students switching between schools, or in behavioral psychology with a twin study where the twins do not wish to share their data with each other. In all these cases, standard practice is to go through user agreements or develop trusted data centers which are allowed to pool data, both of which take lengthy periods of time. Sometimes it is not even possible to establish such a trusted party. Alternatively, estimates may be obtained through algorithms that pool information across the databases to produce combined statistics or estimate models without sharing the data. When privacy is a concern, doing this ``securely" implies two elements: first that the private inputs, the data, from each database are not shared or revealed and second that the estimates produced are accurate to what would be produced if all the data were combined. While guaranteeing the data are not shared, different algorithms can provide different levels of security based on the potential for data leakage through the sharing of intermediate values.

Motivating this work is a real model based on research data that came from multiple sources and could only be combined through a lengthy process with a trusted research network. Specifically we utilized data from a collaboration between the University of Oklahoma Health Sciences Center (OUHSC), the Oklahoma Department of Human Services (OKDHS), and the North Oklahoma County Mental Health Center (NorthCare). Because of the partnership between OUHSC, OKDHS, and NorthCare all data could be gathered and merged into a single data table for analyses, but establishing trust and data sharing agreements between these kinds of agencies in many states is often difficult and tenuous at best. Though the government and university institutions involved in the data collection have been collaborating for over twenty years, the data sharing agreement still took months to establish.

Similar agreements often never come to fruition because sufficient trust and legal protection cannot be forged to release potentially sensitive and identifying information to outside institutions.  Moreover, laws (e.g., the Health Insurance Portability and Accountability Act and the Family Educational Rights and Privacy Act) sometimes preclude data transfers without special permissions from the individuals themselves.  Even when data sharing is mandated by funding agencies or journals, the compliance rate is typically near zero percent; \cite{savage2009empirical} show an example using the PLoS journals.  However, without gathering all the data into a single location, only aggregated summaries would typically be possible and no relationships across variables stored in separate places could be investigated.  In particular, no statistical model could be built from data stored at separate locations without at some point bringing all the data to the same location. Our work shows that for a wide class of models results can be reproduced accurately without needing to physically combine the data, providing a mechanism to allow statistical model building across multiple separate data sources and obviating the need for many data sharing agreements and dramatically reducing the amount of trust required for institutions to collaborate.

\begin{figure}[!ht]
\centering
      \includegraphics[width=0.6\textwidth]{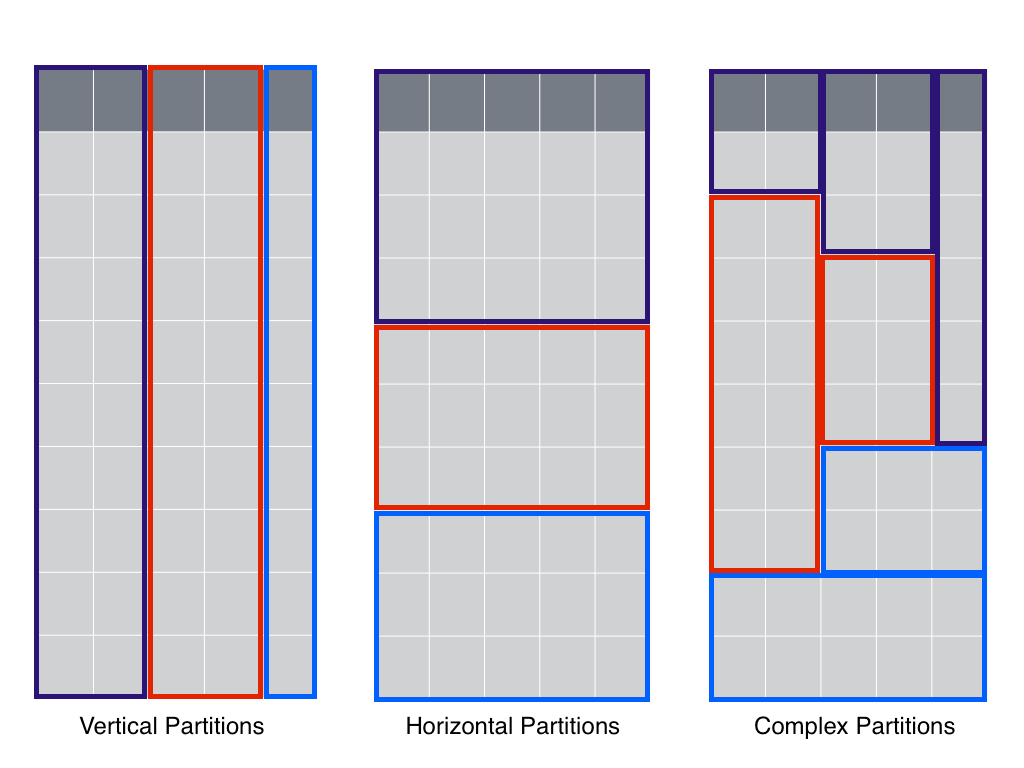}
      \caption{Data partition types, with rows for individual entries and columns for attributes.}\label{fig1}
\end{figure}

The concept known as secure multiparty computation (SMPC) was introduced and originally termed such by \cite{yao1982protocols} (see also \cite{goldwasser1997multi,lindell2009secure}). Various algorithms using SMPC have been proposed in the statistical disclosure control (SDC) literature for the purpose of statistical inference. While much of the SDC work has focused on releasing noisy microdata sets through perturbation, suppression, or generation of synthetic data (see \cite{hundepool2012statistical,WandDeW2001,fienberg2011data,RRR_2003} for more on these methods), the work based on SMPC has focused on providing accurate model estimates without sharing the data and minimizing sharing of intermediate statistics. This is similiar to a secure query, that returns only the answer, on multiple physically seperate data sets.

Partitioned data can be classified as vertically, horizontally, or complexly partitioned, see Figure \ref{fig1}. Vertical partitions imply each database holds the same set of individuals but different variables. The converse, horizontal partitions, implies common variables across databases but different sets of individuals. Finally complex partitions imply some combination of vertical and horizontal. 

Previously in the SMPC literature concerned with parameter estimation, \cite{karr2007secure,lin2010privacy} proposed distributed likelihood estimation (DLE) for horizontally and vertically partitioned data using secure sums and oblivious transfer, \cite{karr2004privacy,karr2009privacy} proposed a secure matrix multiplication algorithm to estimate the sample data covariance matrix without combining data, \cite{ghosh2007secure} showed adaptive regression splines for horizontal data, \cite{sanil2004privacy,karr2005secure}, and \cite{fienberg2006secure,fienberg2009valid,nardi2012achieving} showed secure logistic regression and log-linear models for categorical data. This is also a large body of literature on achieving data mining or machine learning results through SMPC, such as \cite{vaidya2004privacy} or \cite{vaidya2008privacy}, but this literature is concerned only with prediction not parameter estimation.

When dealing with horizontally partitioned data, the solution can often be reached easily, but it is more tricky with vertically or complexly partitioned data. Previous work related to ours focused on two strategies for vertical partitions, either devising secure methods to compute the sample covaraince matrix, \cite{karr2009privacy}, or maximizing over parameters in the case of logistic regression, \cite{nardi2012achieving}. In this paper we will not compute the sample covariance in closed form but achieve it by maximum likelihood.

Commonly SMPC relies on the assumption of semi-honesty, which means that all parties will follow the algorithm as stated and will not collude to try to uncover a third party's information. Sometimes termed ``honest but curious", coined by \cite{kissner2005privacy}, it allows that parties may use available information to try to uncover other parties' data. Another assumption is that the joint answer itself is not risky, such as an estimated covariance matrix in the case of vertically partitioned regression, see \cite{karr2009privacy}. We also make that assumption here, while noting that this may very well not always be the case. Further work should consider methods to combine our algorithm with efforts to minimize any risk of releasing model estimates, but here we focus on getting accurate estimates without sharing data or intermediate statistics, minimizing any possible data leakage.

We present a secure algorithm to calculate the correct multivariate normal log-likelihood for a set of parameters (mean and covariance matrix) given the partitioned data, which can be used with standard optimization techniques to produce maximum likelihood estimates (MLEs) of the parameters. This extends from our previous non-secure algorithm given in \cite{snoke2016accurate}, where it was shown it is possible to accurately compute multivariate normal maximum likelihood estimates without passing any data or statistics that allow for immediate reconstruction of the data. In this paper we go a step further and show that by using secure multiparty computation techniques we can get accurate estimates without sharing any true information between partitions, minimizing any possible data leakage. From a privacy standpoint, this greatly reduces the risk of implementing this algorithm, leaving only the risk from disclosure due to releasing final model estimates.

From the inference perspective, the multivariate normal parameter MLEs are commonly used to estimate a variety of models, such as linear models, factor analysis, principal components analysis (PCA), or structural equation models (SEMs). This approach provides a more general modeling framework from previous work, which focused on specific categories of models. Under our algorithm any analyses relying on the multivariate normal MLE can be performed across partitioned data sources. Additionally, our algorithm naturally extends to complex partitions (see Figure \ref{fig1}), which we demonstrate, while previous methods have focused on horizontal or vertical partitions solely (only offering only hints of solutions for complex partitions). Finally, our method reduces the remaining privacy risks in the vertical partition setting over previous methods to directly computed the sample covariance matrix, and the risk under our algorithm does not increase with the number of partitions, as is the case for some previous methods which we will discuss further in section \ref{sec:prevmeth}.

The rest of this paper is organized as follows. Section 2 reviews related prior methods and discusses potential security or computational improvements. Section 3 covers the mechanics of distributed likelihood estimation. Section 4 presents a detailed explanation of our secure algorithm. Section 5 gives an analysis using real data, showing accurate results without combining the data. Section 6 presents extended simulations to test the computation and accuracy of the algorithm. Section 7 gives final remarks and discussion.

%% file: previous-secure-methods.tex
Here we detail some of the relevant related methods used for secure MLE and covariance modeling, and we cover some of the ways in which we plan to improve on these methods. We focus on previous algorithms that allow for general covariance modeling to show how we improve on them with respect to data leakage. We do not cover specific secure methods such as linear or logistic regression, since our method generalizes to a larger class of models. We also focus on the vertical (and by extension complex) partition case, since MLE in the purely horizontal case has been shown to be easily generalizable (see e.g. \cite{karr2007secure}).

\subsection{Sample covariance estimation by secure matrix multiplication}
\label{sec:smm}
One of the straight-forward applications of secure protocols with statistical quantities is an algorithm for calculating the sample covariance by secure matrix multiplication. The computation of the sample covariance in this way allows for various model estimation, most notably linear and logistic regression as shown in \cite{karr2009privacy,fienberg2009valid,slavkovic2007secure}, and it is a general estimation algorithm, similar to our method.

Suppose we have two parties, with data $X_1 \in \mathbb{R}^{n \times p_1}$ and $X_2 \in \mathbb{R}^{n \times p_2}$, who wish to securely compute the off-diagonal elements of the covariance matrix $X_1^TX_2$. The following algorithm provides such a secure computation.

\begin{algorithm}
\caption{Secure Matrix Multiplication}
\label{alg:smm}
\begin{algorithmic}[1]
\Input $X_1 \in \mathbb{R}^{n \times p_1}$ and $X_2 \in \mathbb{R}^{n \times p_2}$ held by party 1 and 2 respectively
\Output $X_1^TX_2$
\State Party 1 generates an orthonormal matrix $Z \in \mathbb{R}^{n \times a}$, such that $Z_i^T{X_1}_j = 0$ $\forall i,j$ (columns)
\State Party 1 sends $Z$ to party 2
\State Party 2 computes $W = (I - ZZ^T)X_2$ where $I \in \mathbb{R}^{n \times n}$ is the identity matrix
\State Party 2 sends $W$ to party 1
\State Party 1 computes $X_1^TW = X_1^T(I - ZZ^T)X_2 = X_1^TX_2$
\State (Optional) Party 1 sends $X_1^TX_2$ to party 2
\end{algorithmic}
\end{algorithm}

The strength of this algorithm is that it is fairly simple and easy to implement. It can also be implemented when only two parties exists, which we will see in Section~\ref{sec:securesum} is not always possible. The downside to this algorithm is that there exists some data leakage, so it is possible the parties involved can learn about the other parties' data. \cite{karr2009privacy} gives preliminary explanation of the data leakage, and \cite{samizo2016msthesis} provides in-depth analysis, but simply put each party will learn a certain number of linearly independent constraints on the other party's data matrix (contingent on the choice of $a$ in step 1). Additionally, if there are more than two parties, the off-diagonal elements will need to be computed for each pair of databases, increasing both the computational burden ($O(n^2)$) and more importantly the data leakage, since every party will learn a certain number of linearly independent constraints on every other parties' data.

\subsection{Maximum likelihood estimation by secure summation and oblivious transfer}
\label{sec:securesum}
In the case of horizontally partitioned data, \cite{karr2007secure} and \cite{lin2010privacy} describe a method to get MLE estimates for general exponential family models using secure summation. This is fairly straightforward method, and to understand it further consider the example using the multivariate normal log-likelihood.
\begin{align}
  \begin{split}\label{ll1}
    \ell(\mu,\Sigma | Z) = \sum^n_{j=1}{\ell_i(\mu_i,\Sigma_i | z_i)}
 \end{split}
\end{align}
where $Z$ is a theoretical combined database with actual partitions $Z = \begin{pmatrix} X_1 & X_2 & X_3 \\ \end{pmatrix}^T$. We can decompose this into three parts:
\begin{align}
  \begin{split}\label{ll2}
    \ell(\mu,\Sigma | Z) = \sum^{n_{x_1}}_{i=1}\ell_i(\mu,\Sigma | x_{1i}) + \sum^{n_{x_2}}_{j=1}\ell_i(\mu, \Sigma | x_{2i})
     + \sum^{n_{x_3}}_{i=1}\ell_i(\mu, \Sigma | x_{3i})
 \end{split}
\end{align}
which we will refer to as $LL_1$, $LL_2$, and $LL_3$ respectively.

We can get the total sum in a secure manner by adding random noise at the beginning and removing this noise only when all intermediate summations have been completed. This guarantees that as long as the noise is adequately large and random, only the final sum will be learnable to all parties. The steps of the algorithm are as follows:

\begin{algorithm}
\caption{Secure Summation}
\label{alg:ssum}
\begin{algorithmic}[1]
\Input $LL_1, LL_2, LL_3$ held by party 1, 2, and 3 respectively
\Output $LL_{1+2+3}$
\State Party 1 computes $\tilde LL_1 = LL_1 + R$ where $R$ is a large random value
\State Party 2 receives $\tilde LL_1$ and computes $\tilde LL_{1+2} = \tilde LL_1 + LL_2$
\State Party 3 receives $\tilde LL_{1+2}$ and computes $\tilde LL_{1+2+3} = \tilde LL_{1+2} + LL_3$
\State Party 1 receives $\tilde LL_{1+2+3}$ and computes $LL_{1+2+3} = \tilde LL_{1+2+3} - R$
\State Party 1 shares $LL_{1+2+3}$ with 2 and 3
\end{algorithmic}
\end{algorithm}

This routine also allows for a clear understanding of collusion. If party 1 were to collude with party 3, they could send $\tilde LL_1$ to party 3. This would allow party 3 to learn $LL_2$, disclosing party 2's information at no cost to parties 1 or 3. Thus in this horizontal setting with secure summation, we must assume no collusion to ensure security, as well as assuming at least three parties.

For vertically partitioned data, \cite{lin2010privacy} offer a method to get MLE estimates for the general exponential family. The steps are replicated below:
\begin{algorithm}
\caption{Exponential family likelihood by oblivious transfer}
\label{alg:expFam}
\begin{algorithmic}[1]
\Input Data matrices $X_1, X_2$ held by parties 1 and 2 respectively
\Output $\hat \theta = \underset{\theta}{arg max} \: a(\theta)^T \: \Sigma_{i = 1}^n t(X_i) - nc(\theta)$.
\State Party 1 generates a vector $W$ of length $s$, one component of which is $X_i^1$, and the other $s - 1$ of which are random, and sends it to party 2
\State Party 2 computes $t(W_1, X_i^2)...t(W_s,X_i^2)$, generates a random value $\epsilon_i$, and calculates $t(W_1, X_i^2)-\epsilon_i...t(W_s,X_i^2)-\epsilon_i$
\State Party 1 obtains $t(X_i^1, X_i^2) - \epsilon_i$ from these using 1 out of $s$ oblivious transfer (\cite{di2000single})
\State Party 1 holds $\Sigma_i[t(X_i^1, X_i^2) - \epsilon_i]$ and party 2 holds $\Sigma_i \epsilon_i$, which add to $\Sigma_{i=1}^n t(X_i^1,X_i^2)$
\end{algorithmic}
\end{algorithm}

This method is nice because it generalizes to the exponential family, but it is not as strong from a risk standpoint. In the first step, in the oblivious transfer, B has a $1/s$ chance of correctly guessing $X_i^1$, so $s$ must be sufficiently high, which can be computationally difficult. Additionally, in the case of the multivariate normal log-likelihood function $t(X_i) = (X_i X_i^2)$, each party learns the sum of the other party's data, which could be highly disclosive in certain situations. Disclosure from sharing of aggregated statistics has long been researched at places such at the Census Bureau, where the concern primarily stemmed from outliers, see \cite{sullivan1992overview}. More recently many works stemming from the computer science literature have shown there are significant risks, particularly in the case of high dimensional data and in the presence of other external sources, e.g. see \cite{dinur2003revealing,homer2008resolving,calandrino2011you}. Such cases, in part have motivated rise of {\em differential privacy (DP)}, a formal framework for defining the worst-case risk, for review see \cite{dwork2008differential}. The DP itself is beyond the scope of this paper, but in Section~\ref{sec:discussion}, we address a potential interplay of DP with our proposed method. 

%% file: distributed-likelihood-estimation.tex
Distributed likelihood estimation (DLE) refers generally to the concept of calculating a likelihood for estimation purposes, i.e. using maximum likelihood, in separate pieces and combining the resulting likelihoods to get a total value. We will refer to it here specifically in the context of partitioned data, such that likelihoods must be calculated at separate databases and combined to get estimates because the data cannot be physically combined. Furthermore, we focus on the situation where maximum likelihood is achieved through an optimization routine, such as gradient descent, and a central party exists who controls no data but facilitates the optimization. The role of this node will be to choose initial parameters, intermediate parameters, and ultimately convergence criterion. The central node does not control any data, but it is possible they are the most interested in the estimates. This can be imagined as a research node where users can request models and receive estimates. Note that this does \emph{not} mean the central party is a trusted party, as has been implemented by \cite{de2014openpds} or \cite{gaye2014datashield}. The central node, along with the data nodes, does \textit{not} learn any true data or intermediate statistics from the other nodes which is a typical case with a trusted third party. 

A strength of our setup is that having a central party allows us to utilize secure summation technologies with only two data-holding parties, since there are still three total parties. This central party also enables the practicality of our algorithm. Previous methods have either utilized a trusted central party to facilitate an implementation of their algorithm, or they have only addressed a practical implementation in general terms. For example, the secure matrix multiplication or secure summation methods given in section \ref{sec:prevmeth} do not specifically address the question of the original of the research interest or model choice. Our central party is not trusted but is responsible for facilitating the model choice and optimization.

\subsection{Modeling by multivariate normal MLE}
\label{sec:mvndle}
We focus on the multivariate normal MLE, a well-known and flexible modeling framework which allows for the estimation of a variety of models such as linear models, factor models, PCA, and SEMs. Assuming a multivariate normal distribution, the goal of MLE is to maximize the likelihood equation:
\begin{align}
  \begin{split}\label{eq:likelihood}
    L(\mu,\Sigma | Z) = \prod^n_{i=1}{L_i(\mu_i,\Sigma_i | z_i)} = \prod^n_{i=1}{(2\pi)^{-\frac{1}{2}p_i}|{\Sigma_i}|^{-\frac{1}{2}}e^{(z_i-\mu_i)\Sigma_i^{-1}(z_i-\mu_i)^T}}
 \end{split}
\end{align}
for mean and covariance parameters $\mu \in \mathbb{R}^{p}$ and $\Sigma \in \mathbb{R}^{p \times p}$ given a data matrix $Z \in \mathbb{R}^{n \times p}$. The log-likelihood gives a summation rather than a product across each data row, allowing for an easy secure summation application: 
\begin{align}
  \begin{split}\label{eq:loglikelihood}
    \ell(\mu,\Sigma | Z) = \sum^n_{i=1}{\ell_i(\mu_i,\Sigma_i | z_i)} = \sum^n_{i=1}{-\frac{1}{2}[p_i * log(2*\pi) + log(|\Sigma_i|) + (z_i-\mu_i)\Sigma_i^{-1}(z_i-\mu_i)^T]}
 \end{split}
\end{align}
We use the formulation that allows the parameters to vary for each data row, i.e., the full-information likelihood (e.g., see \cite{arbuckle1996full}) dealing with missing data. The log-likelihood will be calculated in separate partitions, and in the vertical case the parameters are partitioned to only pertain to certain variables in the data matrix (see Section \ref{sec:vertdle} for more details). 
Maximizing over equation (\ref{eq:loglikelihood}) gives the MLE estimates $\hat\mu$ and $\hat\Sigma$. With the full data matrix, the maximum can be found in closed form, but in our case because the data are partitioned we can find the maximum by using any standard optimization routine.

\subsection{Horiztonally distributed likelihood}
\label{sec:horzdle}
\begin{figure}[!ht]
\centering
      \includegraphics[width=0.8\textwidth]{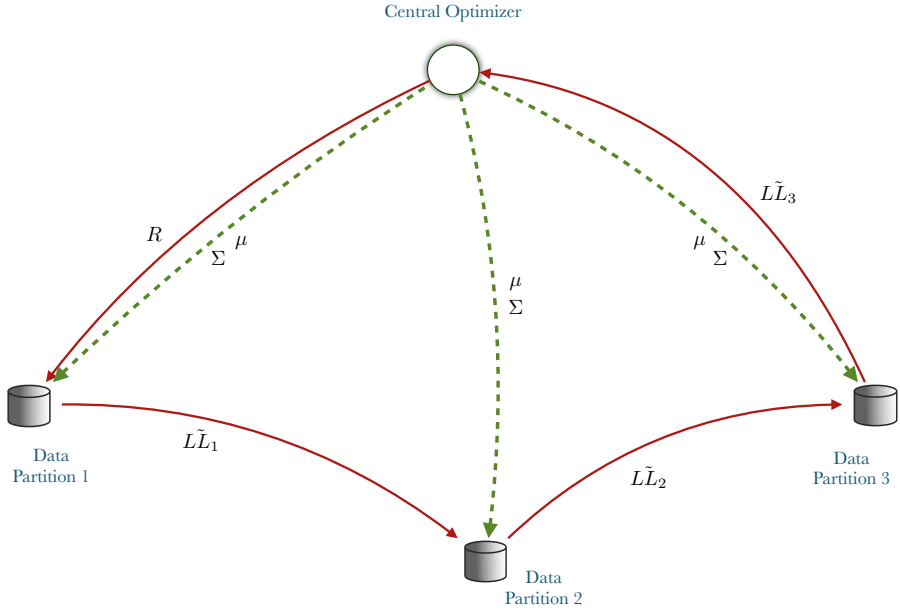}
      \caption{Example of three node system using secure summation for horizontally partitioned data. Red solid lines denote path of noisy log-likelihood, green dashed lines denote path of parameters.}\label{fig:x1}
\end{figure}

Horizontally partitioned data are separated by observations, such that each database has different sets of individuals all measured on the same variables. The parameters are the same for each partition, so DLE requires only a secure summation of total log-likelihoods from each partition as described in section \ref{sec:securesum}. If we consider the log-likelihood given in equation \ref{loglike1}, we can separate this into different elements based on the rows in each partitioned database. All variables are the same across databases, so the parameters relating to the data are the same across each. This is shown in equation \ref{loglike2} for a theoretical combined database $Z$ with partitions $\begin{pmatrix} X_1 & X_2 & X_3 \\ \end{pmatrix}^T$. Note that these partitions do not need to be of equal size.
\begin{align}
  \begin{split}\label{loglike1}
    \ell(\mu,\Sigma | Z) = \sum^n_{i=1}{\ell_i(\mu_i,\Sigma_i | z_i)}
 \end{split}
\end{align}
\begin{align}
  \begin{split}\label{loglike2}
    \ell(\mu,\Sigma | Z) = \sum^{n_{x_1}}_{i=1}\ell_i(\mu,\Sigma | x_{1i}) + \sum^{n_{x_2}}_{i=1}\ell_i(\mu, \Sigma | x_{2i})
     + \sum^{n_{x_3}}_{i=1}\ell_i(\mu, \Sigma | x_{3i})
 \end{split}
\end{align}
In this case, each node receives the full set of parameters from the central optimizer and calculates their portion of the total log-likelihood. The results are added using a secure summation and new parameter estimates are chosen for the next step of the optimization; see Figure \ref{fig:x1} for a visual depiction. In the horizontal case, as noted by \cite{karr2007secure}, the form of the likelihood function does not matter, so applications are not constrained to the multivariate normal case.

\subsection{Vertically distributed likelihood}
\label{sec:vertdle}
Vertically partitioned data are separated by variables, with each database containing different measurements on the same set of individuals. This implies the set of parameters differs at each partition, so as shown in \cite{snoke2016accurate} DLE must combine log-likelihoods calculated using marginal and conditional parameters. (Note that complex partitions can be formulated as a combination of horizontal and vertical DLE, discussed further in section \ref{sec:algorithm}.) Equation (\ref{ll3}) shows the decomposition of the log-likelihood function in the vertical case where $Z = \begin{pmatrix} X_1 & X_2 & X_3 \end{pmatrix}$. Note that these partitions do not need to be of equal size. This adds complexity because some intermediate parameters and statistics must be calculated and shared in order to accurately calculate the total log-likelihood. These must be shared securely, but a simple secure summation of the log-likelihoods is not possible because each partition needs different parameters. Next, we propose how to address this problem. 
\begin{align}
  \begin{split}\label{ll3}
    \ell(\mu,\Sigma | Z) = \sum^n_{i=1}\ell_i(\mu_{x_1},\Sigma_{x_1} | x_{1i}) + \sum^n_{i=1}\ell_i(\hat\mu_{x_2|x_1},\Sigma_{x_2|x_1} | x_{2i}) + \sum^n_{i=1}\ell_i(\hat\mu_{x_3|x_2,x_1},\Sigma_{x_3|x_2,x_1} | x_{3i})
 \end{split}
\end{align}

\subsubsection{Marginal and Conditional Parameters for Vertical Partitions}
\label{sec:schur}
To calculate the partitioned log-likelihoods, we need to calculate marginal and conditional parameters based on $\mu$ and $\Sigma$. Subsetting the parameters as shown in equation \ref{eq7} for the combined dataset, we can write the conditional parameters as shown in Equations \ref{eq9.1}, \ref{eq9.2}, \ref{eq10.1}, and \ref{eq10.2} These relationships come from the well-known properties of the multivariate normal distribution and are also known as the Schur Complement in matrix theory (e.g., see \cite{schur1905neue} and \cite{haynsworth1968schur}).
\begin{align}
  \begin{split}\label{eq7}
  \Sigma = \begin{pmatrix}
            \Sigma_{x_1x_1} & \Sigma_{x_2x_1} & \Sigma_{x_3x_1} \\
            \Sigma_{x_1x_2} & \Sigma_{x_2x_2} & \Sigma_{x_3x_2} \\
            \Sigma_{x_1x_3} & \Sigma_{x_2x_3} & \Sigma_{x_3x_3}
           \end{pmatrix}
  \end{split}
\end{align}  
\begin{align}
  \begin{split}\label{eq8}       
            \mu = \begin{pmatrix}
            \mu_{x_1} & \mu_{x_2} & \mu_{x_3} \\
            \end{pmatrix}
  \end{split}
\end{align}
$\Sigma_{x_kx_k}$ and $\Sigma_{x_kx_l}$ are the model-implied marginal covariance elements for the variables in $X_k$ and covariances between the variables in $X_k$ and $X_l$ respectively. $\mu_{x_k}$ are the mean parameters for the variables in $X_k$.

It is important to note here that these mean and covariance parameters are not the observed sample estimates from the data, since it would be impossible to calculate $\Sigma_{x_kx_l}$ across partitioned data without sharing data. They are parameters defined by our model with values chosen at each step according to the optimization routine. The exception, though, are the conditional mean parameters that are estimated using real data (see Equations \ref{eq9.2} and \ref{eq10.2}), and thus are denoted with the hat notation. Based on these equations, we can estimate a distributed log-likelihood for vertical partitions that is equivalent to the joint log-likelihood.

Note on notation: let $x^+$ denote all variables following $x$ in the node sequence, i.e $x_1^+ = (x_2, x_3)$ (variables in node 2 and 3 for a 3-node system) and $x_2^+ = (x_3)$. In the same way let $x^-$ denote all prior variables. \\

Condition $x_1^+$ on $x_1$:
\begin{align}
  \begin{split}\label{eq9.1}
  \Sigma_{x_1^+|x_1} = \Sigma_{x_1^+x_1^+} - \Sigma_{x_1x_1^+}\Sigma_{x_1x_1}^{-1}\Sigma_{x_1^+x_1}
  = \begin{pmatrix}
            \Sigma_{x_2x_2|x} & \Sigma_{x_2^+x_2|x_1} \\
            \Sigma_{x_2x_2^+|x_1} & \Sigma_{x_2^+x_2^+|x_1}
           \end{pmatrix}
  \end{split}
\end{align} 

\begin{align}
  \begin{split}\label{eq9.2}
  \hat\mu_{x_1^+|x_1} = \mu_{x_1^+} + \Sigma_{x_1x_1^+}\Sigma_{x_1x_1}^{-1}(x_1 - \mu_{x_1})
  = \begin{pmatrix}
            \hat\mu_{x_2|x_1} & \hat\mu_{x_2^+|x_1} \\
            \end{pmatrix}
  \end{split}
\end{align} 

Condition $x_2^+$ on $x_1,x_2$:
\begin{align}
  \begin{split}\label{eq10.1}
  \Sigma_{x_2^+|x_3^-} = \Sigma_{x_2^+x_2^+|x_1} - \Sigma_{x_2x_2^+|x_1}\Sigma_{x_2x_2|x_1}^{-1}\Sigma_{x_2^+x_2|x_1} 
            = \begin{pmatrix}
            \Sigma_{x_3x_3|x_3^-} & \Sigma_{x_3^+x_3|x_3^-} \\
            \Sigma_{x_3x_3^+|x_3^-} & \Sigma_{x_3^+x_3^+|x_3^-}
           \end{pmatrix}
  \end{split}
\end{align} 

\begin{align}
  \begin{split}\label{eq10.2}
  \hat\mu_{x_2^+|x_3^-} = \hat\mu_{x_2^+|x_1} + \Sigma_{x_2x_2^+|x_1}\Sigma_{x_2x_2|x_1}^{-1}(x_2 - \hat\mu_{x_2|x_1}) 
  = \begin{pmatrix}
            \hat\mu_{x_3|x_3^-} & \hat\mu_{x_3^+|x_3^-} \\
            \end{pmatrix}
  \end{split}
\end{align} 

This conditioning process can be repeated for as many partitions as necessary.

%% file: secure-algorithm.tex
Here we describe in more detail our proposed algorithm for secure multiparty log-likelihood estimation with for vertically partitioned data, that easily extends to complex partitions. 
 
\subsection{Notation}
The following terms used in the algorithm and their corresponding equations are, for nodes $k \in {1,...,K}$ with data $\begin{pmatrix} X_1 & X_2 & ... & X_K \\ \end{pmatrix}$ with dimensions $(n \times p_1), (n \times p_2), ..., (n \times p_K)$.
\begin{itemize}
  \item $P_k$, $R_k$, $Q_k$, $M_k$ random noise matrices with dimensions $(n \times p_k), (n \times p_k), (p_k \times n), (n \times p_k)$ respectively
  \item $A_k^1 = \Sigma_{x_kx_k|x_k^-}^{-1}(x_k - \tilde\mu_{x_k|x_k^-} + R_k)$
  \item $A_k^2 = \Sigma_{x_kx_k|x_k^-}^{-1}(x_k - \tilde\mu_{x_k|x_k^-} - R_k) + Q_k$
  \item $B_k = \tilde\mu_{x_k^+|x_k^-} + \Sigma_{x_kx_k^+|x_k^-}A^1_k$
  \item $C_k = \Sigma_{x_kx_k^+|x_k^-}(\Sigma_{x_kx_k|x_k^-})^{-1}$
  \item $\tilde\mu_{x_{k}^+|x_{k+1}^-} = \begin{pmatrix} \tilde\mu_{x_{k+1}|x_{k+1}^-} & \tilde\mu_{x_{k+2}|x_{k+1}^-} & ... & \tilde\mu_{x_{K}|x_{k+1}^-} \end{pmatrix} = B_{k} - M_k - C_{k}(R_{k} - P_{k})$
\end{itemize}

The $+$ and $-$ notation denote all variables in partitions following or preceeding a node, e.g. $\mu_{x_{1}^+|x_{2}^-}$ denotes the conditional mean parameters for all variables in node 2 and following, conditional on all variables in node 1. The $\sim$ and $*$ notation denote noisy versions of the true statistics. There are two noise notations, since the same statistics will have multiple stages of noise addition or removal.

\subsection{Adaptation from non-secure algorithm}
\label{sec:nonsecure}
From \cite{snoke2016accurate}, Algorithm \ref{alg:nontrusted} (shown in appendix \ref{sec:app2}) gives a method for estimating the joint likelihood without sharing any data, but does not provide that nothing can be learned from the statistics which are passed.

There are four elements that we need to be computed and shared in this original algorithm order to obtain correct parameter estimates that we consider to be risky. Two are statistics of the data and two converge to statistics (recall that we use optimization to find maximum likelihood estimates of $\Sigma$ and $\mu$). For each node $k$ they are:
\begin{enumerate}
\item $\Sigma_{x_kx_k| x_k^-}$
\item $\hat\mu_{x_k|x_k^-}$
\item $\Sigma_{x_kx_k^+|x_k^-}$
\item $LL_k$
\end{enumerate}

We handle these elements as follows. First, rather than the central node sending $\Sigma$ and $\mu$ to the first data node and each node calculating the following conditional covariance parameters, the central node will calculate $\Sigma_{x_kx_k| x_k^-}$ for each node $k$. These will be distributed only to each node with the corresponding variables, so nodes will not learn the covariance parameters for other nodes' variables.

Secondly, as mentioned the central node no longer sends $\mu$ to the first node for each node to do the conditioning. The true conditional means are risky even for data nodes containing the corresponding variables to see because they are statistics that depend on the previous nodes' data. To protect this information, we both add noise to $\mu$, and because the conditional means are calculated using unshareable information contained by data nodes and the central node, we must calculate them in parts and combine the pieces in such a way that true values are not recoverable. 

Next, we protect the off-diagonal covariance elements by not passing them unless combined with other terms which cannot be differenced out. Though the off-diagonal elements are not statistics, it is possible through the iteration of the optimization routine that a partition could learn how another partition's data covariance with their own data. Since their own data is known, they may disclose information about the other partition's data.

Lastly we share noisy log-likelihood totals instead of the true values, since these are statistics of data. There are two noise components to the log-likelihood. First, by adding noise to the mean parameters this introduces noise into the log-likelihoods. Second, we multiply by another noise factor that lets us difference out the noise but only at the final step.


We can think of the secure algorithm as having two primary functions, one being a secure summation of log-likelihoods across partitions, and the second being multi-party computation of the conditional mean parameters for each node. Vitally, the solution to these goals work together both in obscuring the true values and in removing the noise by the end in order to obtain the correct joint likelihood value.

\subsection{Algorithm walk-through}
\label{sec:algorithm}
Algorithm \ref{alg:fullAlg} gives the complete $K$ data partition secure process for calculating the total log-likelihood across vertical partitions. There are a number of internal algorithms referenced, given in appendix \ref{sec:app1} and named as shown. We will refer to them by number in the following description of the algorithm.

\begin{algorithm}
\caption{Secure Multiparty Log-Likelihood Estimation for Vertical Partitions}
\label{alg:fullAlg}
\begin{algorithmic}[1]
\Input $\mu \in \mathbb{R}^{p}, \Sigma \in \mathbb{R}^{p \times p}$ (cental node), $X_k \in \mathbb{R}^{n \times p_k}$ $\forall k \in K$ (data nodes)
\Output $LL_{\Sigma_{j = 1}^{K} j}$
\State CN compute: $CN\_Initiate(\mu, \Sigma)$
\State CN $\to$ $DN_1$: $\Sigma_{x_1x_1}, \tilde\mu_{x_1}, P_K$
\State $DN_1$ compute: $EN\_Compute(\tilde\mu_{x_1}, \Sigma_{x_1x_1}, X_1, \emptyset)$
\State $DN_1$ $\to$ CN: $A_1^1, A_1^2$
\State $DN_1$ $\to$ $DN_2$: $\tilde LL_{1}, R_1, Q_1$
\For{k in 2,..., K}
  \State CN compute: $CN\_Adjust(A_{k - 1}^1, \tilde\mu_{x_{k - 1}^+|x_{k - 1}^-}, \Sigma_{x_{k - 1}x_{k - 1}|x_{k - 1}^-})$
  \State CN $\to$ $DN_k$: $\Sigma_{x_kx_k|x_k^-}, B_{k - 1}, C_{k - 1}, P_{k - 1}$
  \State $DN_k$ compute: $EN\_Adjust(B_{k - 1}, C_{k - 1}, R_{k - 1}, P_{k - 1}, Q_{k - 1}, \tilde LL_{\Sigma_{j = 1}^{{k - 1}} j}, M_{k-1})$
  \State $DN_k$ compute: $EN\_Compute(\tilde\mu_{x_k|x_k^-}, \Sigma_{x_kx_k|x_k^-}, X_k, \tilde LL^*_{\Sigma_{j = 1}^{{k - 1}} j})$
  \State $DN_k$ $\to$ CN: $A_k^1, A_k^2$
  \If{k != K}
    \State $DN_k$ $\to$ CN: $\tilde\mu^*_{x_k^+|x_k^-}$
    \State $DN_k$ $\to$ $DN_{k + 1}$: $\tilde LL_{\Sigma_{j = 1}^{{k}} j}, R_k, Q_k, M_k$
  \EndIf
\EndFor
\State $DN_K$ $\to$ $DN_1$: $\tilde LL_{\Sigma_{j = 1}^{{k}} j}, Q_k$
\State $DN_1$ compute: $FN\_Adjust(\tilde LL_{\Sigma_{j = 1}^{K} j}, P_k, Q_k)$
\State $DN_1$ $\to$ CN: $\tilde LL_{\Sigma_{j = 1}^{K} j}^*$
\State CN compute: $CN\_Final(\tilde LL_{\Sigma_{j = 1}^{K} j}^*, P_k, A_k^1, A_k^2, \Sigma_{x_kx_k|x_k^-}$ $\forall k \in K)$
\end{algorithmic}
\end{algorithm}

This algorithm represents one step in an optimization routine, returning the total log-likelihood value for a given set of parameters which will then be used to choose new values under convergence. The inputs are parameters, $\mu$ and $\Sigma$, according to an assumed multivariate normal distribution. The output is a single value, the sum of log-likelihoods across the entire set of partitioned databases.

In the initial step of the algorithm the central node, which runs the optimization and has no data, partitions the parameters corresponding to the variables present at each data node. It produces the marginal and conditional covariance matricies for each partition, which it can do without any information from the data nodes. It also calculates the marginal mean vectors for each node, adding noise to them:
\begin{gather}
  \tilde\mu = \begin{pmatrix} \tilde\mu_{x_1} & \tilde\mu_{x_2} & ... & \tilde\mu_{x_K} \end{pmatrix} = \begin{pmatrix} \mu_{x_1} + P_1 & \mu_{x_2} + P_2 & ... & \mu_{x_K} + P_K \end{pmatrix}
\end{gather}
and saving the noise vectors. It passes on to the first data node the marginal parameters for that node as well as the noise vector which was added to the part of the mean vector pertaining to the $K$-nodes variables. This will be used in a later step in de-noising.

The first data node uses the covariance and noisy mean parameters to calculate a noisy version of the total of the log-likelihood for its data using the following formula:
\begin{gather}
  \tilde LL_{1} = \Sigma^n_{i = 1} [p_1log(2\pi) + log(|\Sigma_{x_1x_1}|) +
  (x_{1i} - \tilde\mu_{x_1} + R_1)\Sigma_{x_1x_1}^{-1}(x_{1i} - \tilde\mu_{x_1} - R_1)^T + R_1^T\Sigma_{x_1x_1}^{-1}R_1]
\end{gather}
with the $R_k$ random noise vectors generated at the node level. See Algorithm \ref{alg:computeLL} for more. This different formula allows the log-likelihood to be calculated with noise that will enable it to be passed securely to other data nodes without revealing the true value, but it is a noise which can be recorded and removed partially at each node and partially at the final step. We can rewrite it as:
\begin{gather}
  \tilde LL_{1} = LL_{1} - \underbrace{A^1_1P_1^T - P_1(A^2_1)^T - P_1\Sigma_{x_1x_1}^{-1}P_1^T}_\text{noise to be removed by central node} + \underbrace{P_1Q_1}_\text{noise to be removed by next data node}
\end{gather}
where there are two noise elements that keep the true values from being revealed. One term will be removed by the following data node, since it cannot be known by the central node. The other terms are removed only at the final step by the central node and serve to keep the following data nodes from revealing the true value. Since the running total is only passed back to the central node at the final step, when it removes all the noise it only learns a single number, the true total log-likelihood across all the data.

This noisy formulation serves a second purpose of calculating the conditional mean parameter, namely through $A^1_k$. As mentioned previously, the conditional mean parameters are statistics, depending on information at both the data node and central node levels, so they must be calculated securely in multiple stages. Because they contains noise generated at the node level, the data nodes can pass $A_k^1$ and $A_k^2$ back to the central node which both facilitates the computation of the conditional mean parameter as well as the eventual de-noising of the log-likelihood.

After performing these computations, the first data node passes the objects discussed back to the central node, and they also pass along the noisy log-likelihood and noise vectors they generated to the next data node. The noisy log-likelihood needs to be passed along to each data node with each noisy total being added before returning to the central node for final de-noising, as in the process of secure summation. The noise vectors passed to the second data node, $R_1$ and $Q_1$, are used both for intermediate de-noising of the log-likeihood, removing noise which cannot be computed by the central node, and in finishing the calculation of the conditional mean parameter. See Algorithm \ref{alg:externalAdjust} for more details.

\begin{figure}[ht]
\centering
      \includegraphics[width=0.8\textwidth]{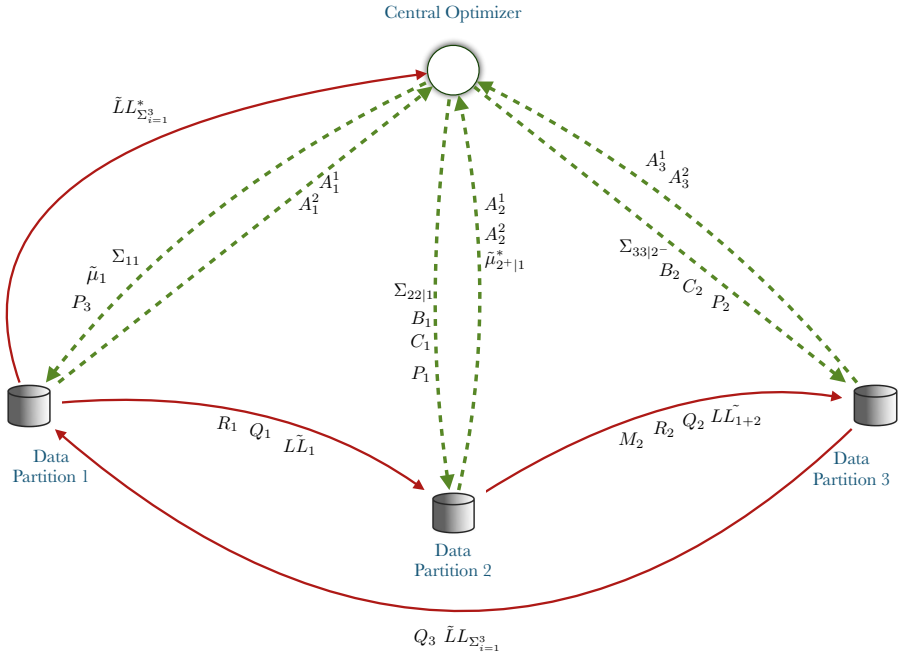}
      \caption{Example three node secure algorithm visual following Algorithm \ref{alg:fullAlg}. Red solid lines denote path of noisy log-likelihood, green dashed lines denote path of parameters.}\label{fig:secureVisual}
\end{figure}

Before moving on to the second data node, the central node needs to make the next step in computing the conditional mean parameter. This combines information received from the previous data node with information controlled by the central node, computing
\begin{gather}
  B_1 = \tilde\mu_{x_1^+} + \Sigma_{x_1x_1^+}A_1^1.
\end{gather}
See Algorithm \ref{alg:centralAdjust} for more details. This is passed on to the next data node.

The second data node performs finalizes the computation of the conditional mean parameter, producing a parameter which is noisy in the same way the original marginal mean parameter ($\tilde\mu_{x_1}$) was noisy. 
\begin{gather}
  \tilde\mu_{x_1^+|x_1} = B_1 - C_1(R_1 - P_1) = \mu_{x_1^+|x_1} + P_{1^+}
\end{gather}
From here the steps start to repeat and look identical at each node. There is one additional wrinkle, as can be seen in Algorithm \ref{alg:externalAdjust}, that from data node two onwards the conditional mean parameters corresponding to the nodes in the following partitions must be passed back to the central node, since the conditioning must stack. Because the central node knows the noise added this parameter, it would be disclosive to return it to the central node. Additional noise $M_k$ is added to prevent this. This noise matrix is also passed along to the next node, so it can be removed in the final step of computing the next conditional mean parameter.

After all the nodes have completed the process of obtaining parameters and calculating noisy log-likelihoods, the total sum is passed back to the first node, not the central node, where it undergoes one final de-noising, see Algorithm \ref{alg:firstAdjust}. This is the same intermediate de-noising that occurs at each data node, and it must occur at the first node before returning to the final node. Finally, the central node receives the noisy value, and knowing all the noise that has accumulated throughout the process is able to remove it and obtain the correct total value, see Algorithm \ref{alg:centralFinal}. Figure \ref{fig:secureVisual} gives a visual representation of the algorithm for a three node system, with the solid red lines denoting the flow of the noisy log-likelihood and the dashed green lines denoting the steps in computing the conditional mean parameters.

\begin{figure}[!ht]
\centering
      \includegraphics[width=0.6\textwidth]{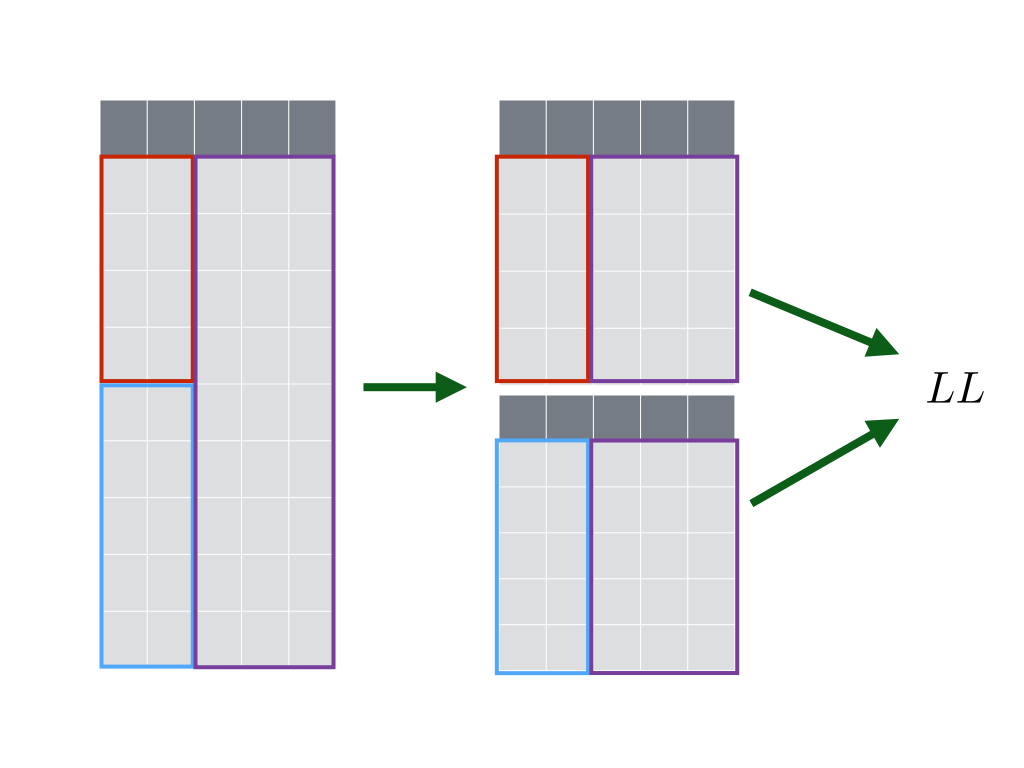}
      \caption{Splitting complex partitions into vertical subroutines}\label{fig:complex_part}
\end{figure}

A simple extension exists for complex partitions by subsetting the routines in horizontally distinct vertical routines and then summing all values at the end. Figure \ref{fig:complex_part} gives a visual depiction of this process. As mentioned earlier, this requires an assumption that the central node knows how the partitions are complexly divided and can appropriately divide the algorithm into subroutines. Each subroutine yields a total log-likelihood value as if it were a pure vertical partition, and each total from all subroutines are added to get the total across all partitions. In the following section we present a real data example, with a complex partition, as a proof of concept for our methods.

%% file: real-data-example.tex

Motivating this paper, we produce model estimates utilizing data collected over four years (2012 - 2015) from multiple sources gathered by a trusted research network between the University of Oklahoma Health Sciences Center (OUHSC), the Oklahoma Department of Human Services (OKDHS), and the North Oklahoma County Mental Health Center (NorthCare). Specifically we examined a study aimed at improving the stability of foster care placements when the foster caregiver was a relative or person known to the family of the child. These are so-call kinship foster placements. In the collaboration families with kinship foster placements were randomized to receive (a) services as usual through OKDHS or (b) OKDHS services and additional services termed Family KINnections from a community resource specialist (CRS) associated with NorthCare.  OUHSC conducted the randomization and OUHSC data collectors gathered baseline data on families randomized to services as usual; however, to aid in subsequent service administration the CRS gathered baseline data for families receiving Famility KINnections. Research is ongoing, but initial work has shown the program reduced the time for a foster family to become certified and increased the stability of a placement by roughly a factor of two. \cite{hunter2016kinship}.

As initially collected, the baseline data (wave 1) were horizontally partitioned. Waves 2 through 7 of data collection were conducted by an OUHSC data collector for both randomization groups, but these later waves were stored in a separate database from the all baseline data, giving us a total of three databases with a complex partition (combination of horizontal and vertical). For the application here we model only the Family Needs Scale (FNS) \cite{fns1988}.  The FNS measures a family's needs for different resources and support with 41 items such as the extent to which the family has a need for ``Having money to pay the bills'' or ``Getting clothes''.  The items are scored from 1 (``Almost Never'') to 5 (``Almost Always'') with higher scores indicating more need.

\begin{table}[ht]
\caption{Simple Latent Growth Curve Model with three estimation methods and three imputation methods. Standard errors given in parenthesis.}\label{tab:real_data}
\centering
\begin{tabular}{cllrrrrrrr}
  && Parameter & Partitioned & Non-partitioned & OpenMx \\
  \hline
  & \vline & $\hat\sigma_{intercept}$ & 0.4468 (0.0236) & 0.4470 (0.0236) & 0.4470 (0.0256) \\ 
  & \vline & $\hat\sigma_{intr\_slp}$ & 0.0000 (NA) & 0.0000 (NA) & 0.0000 (0.0104) \\ 
  Marginal & \vline & $\hat\sigma_{slope}$ & 0.0460 (0.0063) & 0.0462 (0.0063) & 0.0462 (0.0065) \\ 
  Imputation & \vline & $\hat\sigma_{e}$ & 0.3658 (0.0074) & 0.3646 (0.0074) & 0.3646 (0.0077) \\ 
  & \vline & $\hat\mu_{intercept}$ & 2.0230 (0.0318) & 2.0722 (0.0327) & 2.0722 (0.0327) \\ 
  & \vline & $\hat\mu_{slope}$ & -0.0809 (0.0059) & -0.0750 (0.0053) & -0.0750 (0.0053) \\
  \hline 
  & \vline & $\hat\sigma_{intercept}$ & 0.5173 (0.0285) & 0.5183 (0.0285) & 0.5184 (0.0294) \\ 
  & \vline & $\hat\sigma_{intr\_slp}$ & 0.0000 (NA) & 0.0000 (NA) & 0.0000 (0.0062) \\ 
  Joint & \vline & $\hat\sigma_{slope}$ & 0.0701 (0.0064) & 0.0704 (0.0064) & 0.0704 (0.0082) \\ 
  Imputation & \vline & $\hat\sigma_{e}$ & 0.3608 (0.0079) & 0.3594 (0.0078) & 0.3594 (0.0079) \\ 
  & \vline & $\hat\mu_{intercept}$ & 2.0963 (0.0359) & 2.1585 (0.0367) & 2.1585 (0.0367) \\ 
  & \vline & $\hat\mu_{slope}$ & -0.1050 (0.0069) & -0.0978 (0.0063) & -0.0978 (0.0063) \\
  \hline 
  & \vline & $\hat\sigma_{intercept}$ &  &  & 0.5941 (0.0421) \\ 
  & \vline & $\hat\sigma_{intr\_slp}$ &  &  & 0.0000 (0.0191) \\ 
  Full Information & \vline & $\hat\sigma_{slope}$ &  &  & 0.0456 (0.0270) \\ 
  Maximum Likelihood & \vline & $\hat\sigma_{e}$ &  &  & 0.4365 (0.0179) \\ 
  & \vline & $\hat\mu_{intercept}$ &  &  & 2.1097 (0.0483) \\ 
  & \vline & $\hat\mu_{slope}$ &  &  & -0.0697 (0.0129) \\ 
  \hline
\end{tabular}
\end{table}

Table \ref{tab:real_data} shows the parameter estimates for a simple latent growth curve modeling FNS over time with intercept and slope factors and a common residual error among the observed variables. Latent growth models (LGMs) are a way to model a longitudinal trajectory within the SEM framework, see \cite{meredith1990latent}. Algebraically we write the model as:
\begin{gather}
FNS_{ij} = intercept_{i} + \lambda_{j} * slope_{i} + e_{i}
\end{gather}
where $i$ denotes the observation and $j$ denotes the wave. $intercept$ and $slope$ denote the latent factors, and $\lambda_{j}$'s are fixed at $\{0, 1, ..., j-1\}$. We are interested in estimating the covariance matrix for the latent factors (3 parameters), the residual error for the observed variables (1 parameter), and the factor means (2 parameters). We assume here the residual error term is fixed across waves and the observed data has mean zero.

We fit the models in three ways to compare parameter estimates. The first is with the secure algorithm across the three partitions. The data was partitioned as follows: the first partition held roughly two thirds (158 out of 244) of the observations for the first wave of FNS, the second partition held the other subset of the first wave, and the third partition held all observations for waves 2-7. This gave the data a complex partition, similar to that depicted in Figure \ref{fig:complex_part}. The second estimation method was with all data combined using the same optimizer (the \textit{optimx()} \cite{nash2011optimx} function in \textit{R}) that the secure algorithm uses, and the third method again used a combined dataset but using the \textit{OpenMx} \cite{neale2016openmx} package in \textit{R} \cite{r2017} which utilizes a different optimizer. The purpose of having both these methods was to compare our algorithm to both a non-partitioned version of our algorithm and standard software.

In addition to the different methods of estimation, we employed three different approaches to handle missing data. As with most real data problems, missingess is an issue, and currently our algorithm only works for complete data. For the secure algorithm to work with missingess, some parties would need to know the missing data pattern of other parties, which we believe constitutes too much of a privacy risk. Future work should consider ways, if any, to get around this issue. To handle the missingness in the real data, we employed imputation as is commonly done in behavioral and social science research, see \cite{schafer1997analysis}. The first method, marginal imputation, reflects the type of imputation that would need to be done if the data were partitioned, since joint imputation requires all the data. For comparison we also employed joint imputation and full-information maximum likelihood (FIML), a method that subsets the mean and covariance when calculating the log-likelihood in order to calculate the total log-likelihood without imputing or performing listwise deletion. Note that only the \textit{OpenMx} software is capable of performing the FIML approach since this keeps the missing data. Future work could also consider a type of E-M approach that combines our secure estimation with a joint imputation in order to perform joint imputation while preserving the data partitions. This extension should follow naturally and would likely improve estimates over the marginal imputation approach.

We see in table \ref{tab:real_data} that for a given imputation method, the parameter estimates and standard errors for the different estimation methods are very similar, generally identical up to two decimal places, and for inferential purposes they are equivalent. This follows from the theory that the estimates should be the same for the partitioned and non-partitioned methods. We do see larger differences across the different imputation methods, particularly FIML versus the imputation methods, which is understandable given the real data has a decent amount of missingness to impute. Sixty-three percent of all data were missing, varying from 28\% at the initial wave to 84\% at the final wave. In terms of inference, the differences from the imputation method far outweigh any differences from using a partitioned versus non-partitioned estimation procedure. On one hand, we show our method produces equivalent values using partitioned estimation versus non-partitioned estimation, but we must use imputation. This means our inference is limited to the estimation based on complete data models. Additionally, only marginal imputation is currently possible, since the data cannot be shared for joint imputation. Further work could consider developing either an E-M approach or a FIML apporach in the partitioned setting to improve estimation when missing data exists. In the following section we ran further simulations without missing data to assess the general numerical accuracy of parameters estimated using the secure algorithm.

%% file: simulations.tex
\begin{figure}[!ht]
\centering
      \includegraphics[width=\textwidth]{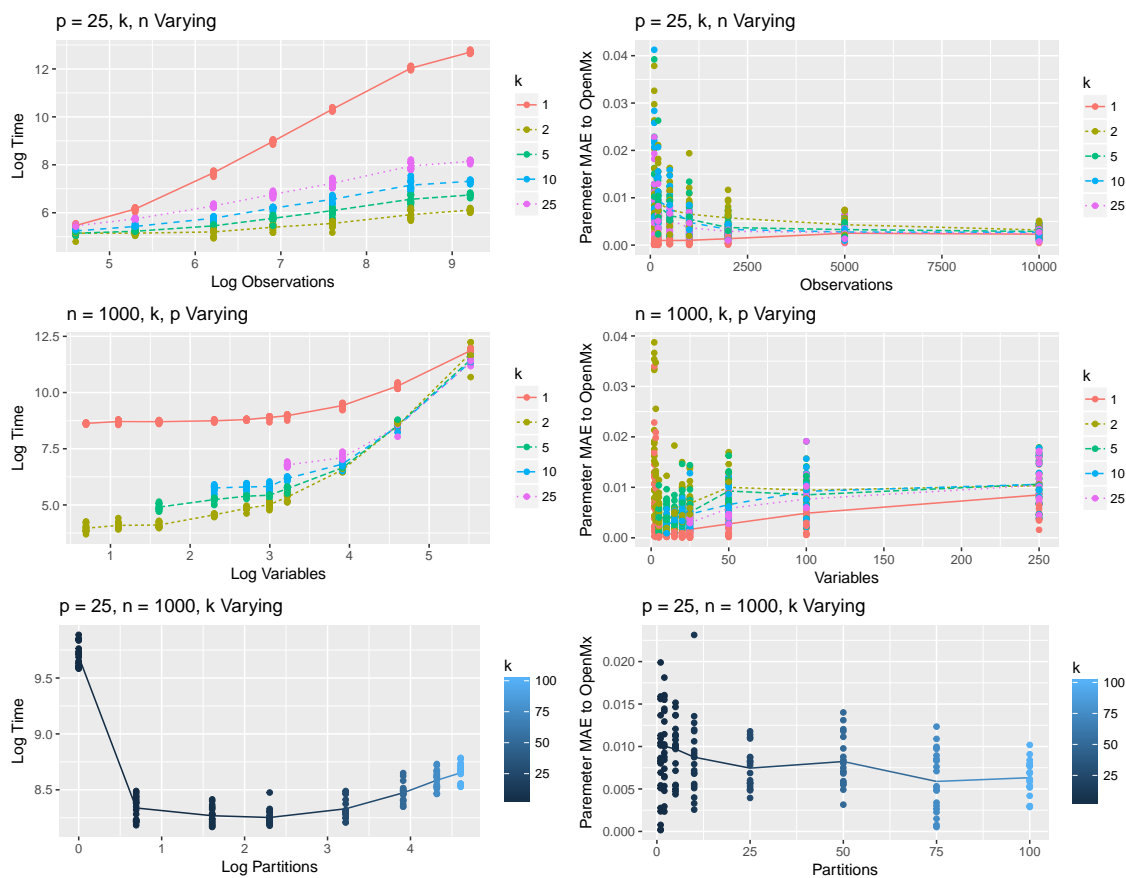}
      \caption{Simulations for varying numbers of variables (p), observations (n), and nodes (k)}\label{fig:simulations}
\end{figure}

We ran simulations (available along with the algorithm at \url{https://github.com/jsnoke/Firewall}) to test the computational complexity of the algorithm as the number of observations, variables, and partitions increase. We expect the algorithm to scale normally with respect to variables and observations, and we are interested to see the results as the number of nodes increase. In one respect, the run time will slow because the number of total computations is increasing, but as each node holds fewer variables the covariance inversions will shrink, reducing run time. As we see from figure \ref{fig:simulations}, this tradeoff does exist. When we vary number of observations, each line for a different number of nodes shows the same slope, but shifted from one another. In the case of varying number of variables, again each line shows the same curve, but the optimal point is in a different place due to the relationship between $p$ and $k$. When we look at fixed $p$ and $n$, we again see an optimal number of nodes ($k = 10$) for $p = 25$ and then a uptick after that.

We also tested the accuracy of the estimates compared to standard software for non-partitioned estimates. Theoretically these should be the identical, but we were curious to investigate possible numerical differences in the actual computation. We see that overall there is less than $0.01$ error, which is less than 1\% error for the scale of the parameters. Recall that theoretically there is no difference between the maximum likelihood estimates from the partitioned or non-partitioned algorithms, so these differences occur due to numerical precision. For practical purposes, the secure partitioned algorithm would produce equivalent results as was exhibited in the real data example.

%% file: discussion.tex
The method presented in this paper represents a new alternative to the current methods for paritioned model estimation. It extends previous work by generalizing the method to any type of covariance modeling, strengthens the privacy guarantees, particularly for increasing numbers of partitions, and allows horiztonal, vertical, or complex partitions within the same algorithm. These theoretical results were confirmed with a real data example, and the accuracy was further tested through simulations.

For practical purposes, the algorithm presented here provides a solution for partitioned estimation when researchers are interested in covariance modeling under a multivariate normal assumption and it guarantees no sharing of data or intermediate statistics for two or data parties. Some practical issues still exist, such as the need to impute missingness marginally and the general complexity of the algorithm, but these are details which can and should be further tuned in future work. The algorithm contains a fair level of complexity, with numerous steps needed to obtain the correct values without sharing any true information. It may be possible in future work to simplify the algorithm, potentially making it easier to understand or save on computation time. By giving detailed algorithms and making the code easily accessible we hope to facilitate any future implementations or improvements of this methodology.

As a final remark, there has been a move in the privacy literature towards formal privacy methods, such as Differential Privacy. While this work guarantees security under a different definition, we believe it future work should look for ways to combine the methods here with formal protections such as offered by perturbations of the output statistics. In some applications it may be the case that revealing the true estimated model (or mean and covariance matrix) constitutes a violation of privacy. For problems where that is not the case, the algorithm presented here provides a strong guarantee of security that multiple parties can engage in partitioned estimation with only the final model values being shared.